%% file: main_revised_v8.tex
\documentclass{article}
\usepackage{spconf}
\usepackage{graphicx}
\usepackage{latexsym}
\usepackage{subfigure}
\usepackage{multirow}
\usepackage{comment}
\usepackage{amsmath}
\usepackage{amsthm}
\usepackage{amssymb}
\usepackage{multirow}
\usepackage{algorithm}
\usepackage{algorithmic}
\usepackage{color}
\usepackage{url}
\usepackage[hidelinks]{hyperref}

\usepackage{microtype}
\usepackage{booktabs} 
\usepackage{cite}

\usepackage{mathtools}

\makeatletter
\renewcommand\section{\@startsection{section}{1}{\z@}{-2.4ex plus -0.8ex minus -.2ex}{1.2ex plus .2ex}{\large\bf}}
\renewcommand\subsection{\@startsection{subsection}{2}{\z@}{-2.0ex plus -0.8ex minus -.2ex}{0.8ex plus .2ex}{\normalsize\bf}}
\makeatother

\title{Multivariate signal restoration via Multifold Graph Learning}
\name{Haruki Yokota and Yuichi Tanaka\thanks{This work was supported by JSPS KAKENHI Grant Number 25KJ1755.}}
\address{Graduate School of Engineering, The University of Osaka, Japan}
\begin{document}
\ninept
\setlength{\abovedisplayskip}{4pt plus 2pt minus 2pt}
\setlength{\belowdisplayskip}{4pt plus 2pt minus 2pt}
\setlength{\abovedisplayshortskip}{0pt plus 2pt}
\setlength{\belowdisplayshortskip}{3pt plus 2pt minus 1pt}
\maketitle
\begin{abstract}
We propose a method for restoring noisy and incomplete multivariate signals by learning multifold graphs. 
Spatial relationships among channels can be utilized to recover missing values and suppress noise. 
However, these relationships can vary across frequency components and observation windows while retaining common connectivity patterns. 
We refer to the graphs associated with individual components and windows as \textit{multifold graphs}. 
We model multifold graphs as nonnegative combinations of shared prototype graphs. 
We formulate signal restoration and multifold graph learning as a joint optimization problem and solve it by alternating optimization.
We unroll the resulting algorithm into a neural network with trainable regularization weights, step sizes, and filter coefficients.
The network is trained in a self-supervised manner.
Synthetic experiments show improved graph estimation accuracy over the compared methods. 
On real-world weather data, the proposed network achieves competitive completion performance with substantially fewer trainable parameters than a self-attention-based model.
\end{abstract}
\section{Introduction}
\label{sec:intro}

Spatial relationships among channels can be utilized to restore noisy or incomplete multivariate signals.
Graphs represent these relationships in sensor, brain, and transportation networks, providing the structure used by graph signal processing~\cite{shuman_emerging_2013,chen_discrete_2015,ortega_graph_2018} and graph neural networks~\cite{kipf_semisupervised_2017,wu_inductive_2021}.
When the underlying graph is unknown, its edge weights are estimated from observed signals by \textit{graph learning}~\cite{dong_learning_2016,kalofolias_how_2016,dong_learning_2019,mateos_connecting_2019}, typically as a single graph for the whole signal.

\begin{figure*}[t]
  \centering
  \includegraphics[width=0.98\textwidth]{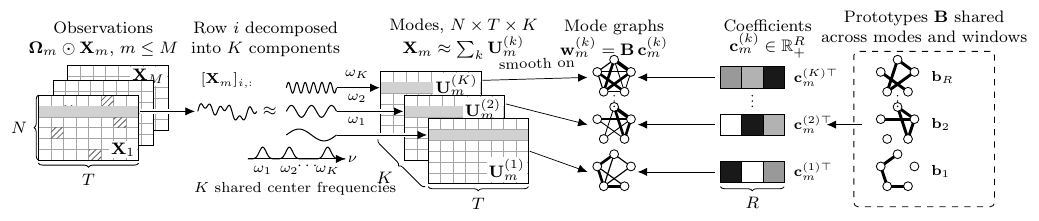}
  \caption{Multifold graph model. Each window is a sum of $K$ narrowband modes, and each mode is smooth on its own graph, a nonnegative mixture of $R$ prototypes shared across modes and windows.}
  \label{fig:model}
\end{figure*}

However, relationships among channels can differ across the frequency components of a multivariate signal.
For example, diurnal sea-level pressure amplitudes differ strongly between coastal and inland stations, whereas semidiurnal amplitudes vary mainly with latitude~\cite{dai_diurnal_1999}.
Thus, stations with similar amplitudes in one component need not be similar in another.
Similarly, neural recordings exhibit frequency-dependent synchronization patterns across brain regions~\cite{klimesch_frequency_2018,gonzalez_lowfrequency_2022}.
Relationships among channels may also vary across observation windows~\cite{kalofolias_learning_2017,natali_learning_2022,yokota_timevarying_2026}.
We refer to the graphs associated with individual frequency components and observation windows as \textit{multifold graphs}.

To restore multivariate signals, we need to estimate the multifold graphs from observations.
However, noisy and incomplete observations may not provide enough information to reliably estimate each of these graphs independently.
Moreover, the graphs provide spatial information for signal restoration, while reliable graph estimation depends on the quality of the recovered signal components. 
Thus, the challenge is to constrain graph estimation without losing component- and window-specific relationships, while jointly restoring the signals.

Existing methods address different aspects of this problem.
Time-varying graph mode decomposition (TVGMD)~\cite{urrehman_timevarying_2023} jointly estimates narrowband temporal modes and their graphs, but does not share graph structure across modes and windows.
The graph-dictionary signal model (GraphDict)~\cite{cappelletti_graphdictionary_2026} represents instantaneous graphs as nonnegative combinations of graph atoms, but does not decompose the signal into frequency components.
Inpainting-driven graph learning (IDGL)~\cite{batreddy_inpainting_2025} jointly learns the graph and completes the signal, but estimates one graph per window without distinguishing frequency components.
These approaches motivate jointly restoring signals and learning component-specific graphs with structure shared across frequency components and observation windows.

In this paper, we propose a method that restores noisy and incomplete multivariate signals by learning multifold graphs.
We assume that each graph can be represented as a nonnegative combination of a few prototype graphs shared across frequency components and observation windows.
This representation constrains the graphs through shared prototypes, while the mixing coefficients capture component- and window-specific differences in connectivity.
Observations from multiple components and windows contribute to estimating the shared prototypes.
We formulate mode decomposition, graph learning, and signal restoration as a joint optimization problem and solve it by alternating optimization~\cite{bertsekas_parallel_1997}.

We further \textit{unroll} the alternating algorithm into a neural network~\cite{gregor_learning_2010,monga_algorithm_2021} whose layers correspond to iterations.
The unrolled network has only a few hundred trainable parameters, corresponding to regularization weights, step sizes, and filter coefficients.
We train the network in a self-supervised manner.

We evaluate the proposed method in terms of graph estimation accuracy and signal completion performance.
On synthetic signals, the unrolled network achieves the lowest graph estimation error among the compared methods.
On real-world temperature data, it achieves the lowest or second-lowest completion error across five missing-data settings while using about 1/1500 as many trainable parameters as a strong neural network baseline for time-series imputation~\cite{du_saits_2023}.

\textit{Notation.}
Vectors and matrices are denoted by bold lowercase and uppercase letters, respectively, $\mathbf{1}$ is the all-ones vector, and $\|\cdot\|_2$ and $\|\cdot\|_F$ denote the Euclidean and Frobenius norms.
A weighted undirected graph on $N$ nodes has $E=N(N-1)/2$ candidate edges $e=(i,j)$, $i<j$, with weights vectorized as $\mathbf{w}\in\mathbb{R}_+^{E}$.
The incidence operator $\mathbf{S}\in\{0,1\}^{N\times E}$ maps edge weights to the degree vector $\mathbf{S}\mathbf{w}$, and $\mathbf{L}(\mathbf{w})$ is the combinatorial Laplacian.
The symbols $\odot$ and $\oslash$ denote elementwise multiplication and division, and exponentiation of a vector is elementwise.

\section{Multifold Graph Model}
\label{sec:method}

We model the signal in each observation window as a sum of narrowband frequency components, each associated with a graph (Fig.~\ref{fig:model}).
We then define temporal bandwidth and graph signal smoothness to characterize these components.

\subsection{Signal and Multifold Graph Model}
\label{sec:problem}

Let $\mathbf{X}_m\in\mathbb{R}^{N\times T}$ denote the $N$-channel signal in window $m=1,\dots,M$, with $T$ samples per window.
We approximate $\mathbf{X}_m$ as $\sum_{k=1}^{K}\mathbf{U}_m^{(k)}$, where $\mathbf{U}_m^{(k)}\in\mathbb{R}^{N\times T}$ is the $k$th narrowband mode.
Its spectral energy is concentrated around a center frequency $\omega_k$, shared across windows.
Modes are indexed by increasing center frequency.

We associate each mode $\mathbf{U}_m^{(k)}$ with a graph whose nodes correspond to the channels and whose edge weights are $\mathbf{w}_m^{(k)}\in\mathbb{R}_+^E$.
We assume that the mode is smooth on this graph.
To capture common connectivity patterns, we represent $\mathbf{w}_m^{(k)}$ using $R$ shared prototype edge weights:
\begin{equation}
  \mathbf{w}_m^{(k)} = \mathbf{B}\,\mathbf{c}_m^{(k)},
  \label{eq:multifold}
\end{equation}
where $\mathbf{B}=[\mathbf{b}_1,\dots,\mathbf{b}_R]\in\mathbb{R}_+^{E\times R}$ collects the prototype edge weights shared across modes and windows.
The nonnegative coefficients $\mathbf{c}_m^{(k)}\in\mathbb{R}_+^R$ determine their contribution.

Let $\boldsymbol{\Omega}_m\in\{0,1\}^{N\times T}$ denote the observation mask, with ones at observed entries and zeros at missing entries.
Given the masked observations $\{\boldsymbol{\Omega}_m\odot\mathbf{X}_m\}_{m=1}^{M}$, we jointly estimate the modes, center frequencies, prototype graphs, and mixing coefficients.

\subsection{Temporal Bandwidth and Graph Signal Smoothness}
\label{sec:regularization}
We first define a temporal bandwidth measure for a mode $\mathbf{U}\in\mathbb{R}^{N\times T}$ with center frequency $\omega$.
For clarity, we omit the mode and window indices.
The bandwidth of $\mathbf{U}$ is expressed using the one-sided discrete Fourier transform (DFT) following the multivariate variational mode decomposition (MVMD)~\cite{urrehman_multivariate_2019}:
\begin{equation}
  \mathrm{BW}(\mathbf{U};\omega)=\sum\nolimits_{\nu\in\mathcal{V}_{+}}\kappa_\nu(\nu-\omega)^{2}\,\|\hat{\mathbf{U}}(\nu)\|_2^{2},
  \label{eq:bw}
\end{equation}
where $\hat{\mathbf{U}}(\nu)=\sum_{t=0}^{T-1}\mathbf{U}_{:,t}e^{-j2\pi\nu t/T}$ is the DFT coefficient for frequency $\nu$.
The set $\mathcal{V}_{+}=\{0,\ldots,\lfloor T/2\rfloor\}$ contains the nonnegative frequency bins.
The frequencies $\nu/T$ and $\omega/T$ are expressed in cycles per sample.
For $\nu\in\mathcal{V}_{+}$, we set $\kappa_\nu=2/T$ when $0<\nu<T/2$ and $\kappa_\nu=1/T$ otherwise.

We next quantify graph signal smoothness using the squared differences between the signals of adjacent nodes.
Let $\mathbf{z}(\mathbf{U})\in\mathbb{R}_+^E$ collect the squared differences $[\mathbf{z}(\mathbf{U})]_e=\|\mathbf{U}_{i,:}-\mathbf{U}_{j,:}\|_2^2$ for each edge $e=(i,j)$.
The graph signal smoothness is $\langle\mathbf{w},\mathbf{z}(\mathbf{U})\rangle=\operatorname{tr}(\mathbf{U}^{\top}\mathbf{L}(\mathbf{w})\mathbf{U})$, where $\langle\cdot,\cdot\rangle$ denotes the Euclidean inner product and $\operatorname{tr}(\cdot)$ the trace.
This quantity is small when strongly connected channels have similar temporal signals.

Hereafter, we consider the normalized graph signal smoothness $\langle\mathbf{w},\mathbf{z}(\mathbf{U})\rangle/\bar z$, where $\bar z>0$ is a fixed factor.

\section{Multifold Graph Learning for Multivariate Signal Restoration}
\label{sec:opt}

In this section, we formulate a joint optimization problem for signal restoration and multifold graph learning.
We then derive an alternating algorithm and unroll it into a neural network.

\subsection{Joint Problem and Block-Coordinate Updates}
\label{sec:formulation}
\label{sec:updates}

To couple signal restoration with mode decomposition, we introduce auxiliary completed signals $\mathbf{Y}_m\in\mathbb{R}^{N\times T}$ for each window.
Let $\mathcal{Q}$ collect the optimization variables: $\{\mathbf{U}_m^{(k)}\}$, $\{\mathbf{Y}_m\}$, $\{\omega_k\}$, $\mathbf{B}$, and $\{\mathbf{c}_m^{(k)}\}$, over all modes and windows.
We jointly estimate these variables by solving
\begin{equation}
\begin{aligned}
  \min_{\mathcal{Q}}&\ \sum\nolimits_{m}\big[\|\boldsymbol{\Omega}_m\odot(\mathbf{X}_m-\mathbf{Y}_m)\|_F^{2}
    +\rho\|\mathbf{Y}_m-\textstyle\sum_{k}\mathbf{U}_m^{(k)}\|_F^{2}\big]\\
  &\!\!+2\rho\widehat\alpha\sum\nolimits_{m,k}\mathrm{BW}(\mathbf{U}_m^{(k)};\omega_k)
   +\tfrac{\lambda}{\bar z}\sum\nolimits_{m,k}\langle\mathbf{B}\mathbf{c}_m^{(k)},\mathbf{z}(\mathbf{U}_m^{(k)})\rangle\\
  &\!\!+\beta_1\sum\nolimits_{m,k}\|\mathbf{B}\mathbf{c}_m^{(k)}\|_2^{2}
   -\beta_2\sum\nolimits_{m,k}\mathbf{1}^{\top}\log(\mathbf{S}\mathbf{B}\mathbf{c}_m^{(k)}).
\end{aligned}
\label{eq:joint}
\end{equation}
Here, $\rho,\alpha,\lambda,\beta_1,\beta_2>0$ control the relative weights of the terms.
We set \(\widehat{\alpha}=\alpha/T^2\) to compensate for the energy of the frequency components.
The objective consists of the following three groups.

\noindent 1) \textit{Signal reconstruction}:
The fidelity term $\|\boldsymbol{\Omega}_m\odot(\mathbf{X}_m-\mathbf{Y}_m)\|_F^2$ fits $\mathbf{Y}_m$ to the observed entries.
The penalty $\|\mathbf{Y}_m-\sum_k\mathbf{U}_m^{(k)}\|_F^2$ keeps $\mathbf{Y}_m$ close to the mode sum.

\noindent 2) \textit{Mode regularization}:
The bandwidth penalty $\mathrm{BW}(\mathbf{U}_m^{(k)};\omega_k)$ promotes narrowband modes.
The inner product $\langle\mathbf{B}\mathbf{c}_m^{(k)},\mathbf{z}(\mathbf{U}_m^{(k)})\rangle$ promotes smoothness on each mode's graph and couples signal estimation with graph learning.

\noindent 3) \textit{Graph regularization}:
The penalty $\|\mathbf{B}\mathbf{c}_m^{(k)}\|_2^2$ controls the magnitudes of the edge weights.
The barrier $-\mathbf{1}^{\top}\log(\mathbf{S}\mathbf{B}\mathbf{c}_m^{(k)})$ prevents isolated nodes~\cite{kalofolias_how_2016}, with elementwise logarithms.
The prototypes and mixing coefficients have a scaling ambiguity, since scaling a prototype up and its coefficients down leaves every mixed graph unchanged. 
To fix the scale, we constrain each prototype to have total edge weight $N/2$, corresponding to unit average degree.
The numbers of modes $K$ and prototypes $R$ are fixed.

The joint problem in \eqref{eq:joint} is nonconvex.
We solve it by alternating between signal reconstruction, frequency updates, mixing-coefficient updates, and prototype updates.
Algorithm~\ref{alg:multifold} summarizes the update sequence.
We perform $J$ outer iterations.
Within each block, we repeat the updates until a predetermined stopping criterion is met or an iteration limit is reached.
We explain each update below.

\vspace{2pt}\noindent\textbf{Signal updates}: 
With the graphs and center frequencies fixed, signal updates are independent across windows. We omit the index $m$ for clarity.
We update each mode using the residual
$\mathbf{R}^{(k)}=\mathbf{Y}-\sum_{i\neq k}\mathbf{U}^{(i)}$,
formed from the latest estimates of the other modes.
The minimizer is the Wiener filter in the frequency domain:
\begin{equation}
  \hat{\mathbf{U}}^{(k)}(\nu)
  =\big[a_k(\nu)\mathbf{I}
  +\tfrac{\lambda}{\rho\bar z}\mathbf{L}^{(k)}\big]^{-1}
  \hat{\mathbf{R}}^{(k)}(\nu),
  \label{eq:wiener}
\end{equation}
where $a_k(\nu)=1+2\widehat{\alpha}(\nu-\omega_k)^2$,
$\mathbf{L}^{(k)}=\mathbf{L}(\mathbf{B}\mathbf{c}^{(k)})$,
and $\hat{\mathbf{R}}^{(k)}(\nu)$ is the DFT of the residual.
After transforming the updated modes back to the time domain, $\mathbf{Y}$ is updated by minimizing the first two terms of \eqref{eq:joint} with fixed modes.
The minimizer is given in closed form:
\begin{equation}
Y_{it}=
\frac{\Omega_{it}X_{it}+\rho\sum_k U_{it}^{(k)}}
{\Omega_{it}+\rho}.
\label{eq:updateY}
\end{equation}

\noindent\textbf{Center-frequency updates}:
With fixed modes, only the bandwidth term in \eqref{eq:joint} depends on $\omega_k$.
Setting its derivative to zero gives
\begin{equation}
\omega_k=
\frac{
\sum_{m=1}^{M}\sum_{\nu\in\mathcal{V}_{+}}
\kappa_\nu\,\nu\,\|\hat{\mathbf{U}}_m^{(k)}(\nu)\|_2^2
}{
\sum_{m=1}^{M}\sum_{\nu\in\mathcal{V}_{+}}
\kappa_\nu\,\|\hat{\mathbf{U}}_m^{(k)}(\nu)\|_2^2
}.
\label{eq:updateOmega}
\end{equation}
This weighted spectral centroid pools energy across all channels and windows.
We retain the previous frequency if the denominator is zero and skip this update when the center frequencies are given a priori.

\noindent\textbf{Mixing-coefficient updates}: 
With the modes and prototypes fixed, each coefficient vector can be updated independently.
For one mode and window, write $\mathbf{c}=\mathbf{c}_m^{(k)}$ and $\mathbf{z}=\mathbf{z}(\mathbf{U}_m^{(k)})$.
Only the graph signal smoothness, squared edge-weight penalty, and logarithmic degree barrier in \eqref{eq:joint} depend on $\mathbf{c}$.
Their combined gradient is $\nabla^{+}-\nabla^{-}$, where $\nabla^{+}=\frac{\lambda}{\bar z}\mathbf{B}^{\top}\mathbf{z}+2\beta_1\mathbf{B}^{\top}\mathbf{B}\mathbf{c}$ and $\nabla^{-}=\beta_2\mathbf{B}^{\top}\mathbf{S}^{\top}\bigl(\mathbf{1}\oslash(\mathbf{S}\mathbf{B}\mathbf{c})\bigr)$.
We apply the multiplicative update~\cite{lee_algorithms_2001}
\begin{equation}
\mathbf{c}\leftarrow\mathbf{c}\odot
\bigl(\nabla^{-}\oslash\nabla^{+}\bigr)^{\eta^{C}},
\qquad 0<\eta^{C}\leq1,
\label{eq:updC}
\end{equation}
where $\eta^{C}$ controls the relaxation of the update.

\noindent\textbf{Prototype updates}: 
With the modes and mixing coefficients fixed, we update the shared prototypes using contributions from all modes and windows.
The gradient with respect to $\mathbf{B}$ is $\mathbf{G}^{+}-\mathbf{G}^{-}$, where
\begin{equation}
\begin{aligned}
\mathbf{G}^{+}
&=\sum_{m,k}
\left[
\frac{\lambda}{\bar z}\mathbf{z}(\mathbf{U}_m^{(k)})
+2\beta_1\mathbf{w}_m^{(k)}
\right]
(\mathbf{c}_m^{(k)})^{\top},\\
\mathbf{G}^{-}
&=\beta_2\sum_{m,k}
\mathbf{S}^{\top}
\left(\mathbf{1}\oslash(\mathbf{S}\mathbf{w}_m^{(k)})\right)
(\mathbf{c}_m^{(k)})^{\top},
\end{aligned}
\label{eq:gradB}
\end{equation}
with $\mathbf{w}_m^{(k)}=\mathbf{B}\mathbf{c}_m^{(k)}$.
We update $\mathbf{B}\leftarrow\mathbf{B}\odot(\mathbf{G}^{-}\oslash\mathbf{G}^{+})^{\eta^B}$, where $0<\eta^B\leq1$.
For each updated prototype, let $s_r=2\mathbf{1}^{\top}\mathbf{b}_r/N$.
We then set $\mathbf{b}_r\leftarrow\mathbf{b}_r/s_r$ and $c_{m,r}^{(k)}\leftarrow s_r c_{m,r}^{(k)}$ for all $m,k$.
This rescaling gives each prototype unit average weighted degree while preserving every graph and the objective value.

Finally, we repeat the signal and frequency updates with the final graphs and sum the modes to reconstruct the signal.

\begin{algorithm}[t]
  \caption{Joint multifold graph learning and signal restoration}
  \label{alg:multifold}
  \renewcommand{\algorithmicrequire}{\textbf{Input:}}
  \renewcommand{\algorithmicensure}{\textbf{Output:}}
  \begin{algorithmic}[1]
    \REQUIRE $\{\boldsymbol{\Omega}_m\odot\mathbf X_m\}$, $\{\boldsymbol{\Omega}_m\}$, $J$, initial values of $\mathcal Q$
    \ENSURE $\{\mathbf X_m^\star\}$, $\{\mathbf w_m^{(k)}\}$
    \FOR{$j=1,\ldots,J$}
      \STATE Iterate: $\mathbf U_m^{(k)}$ by \eqref{eq:wiener} for all $k,m$, $\mathbf Y_m$ by completion, $\omega_k$ by the centroid unless fixed
      \STATE Iterate: $\mathbf c_m^{(k)}$ by \eqref{eq:updC} for all $k,m$
      \STATE Iterate: $\mathbf B$ by the multiplicative update, then normalize
    \ENDFOR
    \STATE Repeat line 2 with the final $\mathbf B$, $\{\mathbf c_m^{(k)}\}$
    \STATE \textbf{return} $\mathbf X_m^\star=\sum_{k}\mathbf U_m^{(k)}$, $\mathbf w_m^{(k)}=\mathbf B\mathbf c_m^{(k)}$
  \end{algorithmic}
\end{algorithm}

\subsection{Unrolled Network}
\label{sec:unroll}

We unroll Algorithm~\ref{alg:multifold} into a neural network with $J$ layers, each corresponding to one outer iteration~\cite{gregor_learning_2010,monga_algorithm_2021}.
Each layer follows the same update sequence, with $S_1$ iterations for the signal and frequency updates and $S_2$ iterations each for the coefficient and prototype updates.
$S_1$ and $S_2$ are fixed, while the regularization weights, relaxation exponents, and filter coefficients are trainable.

The mode update in \eqref{eq:wiener} requires repeatedly solving graph-dependent linear systems.
For the unrolled network, we replace the inverse operator with a learnable polynomial surrogate of degree $P$ and trainable coefficients $\theta_{k,p}$ for each mode:
\begin{equation}
  \hat{\mathbf{U}}^{(k)}(\nu)
  =\sum_{p=0}^{P}\theta_{k,p}\,
  \frac{\big(\tilde{\mathbf{L}}^{(k)}\big)^p
  \hat{\mathbf{R}}^{(k)}(\nu)}{a_k(\nu)^{p+1}}.
  \label{eq:neumann}
\end{equation}
$\tilde{\mathbf L}^{(k)}=\tfrac12(\mathbf D^{(k)})^{-1/2}\mathbf L^{(k)}(\mathbf D^{(k)})^{-1/2}$ is the normalized graph Laplacian defined with $\mathbf D^{(k)}=\operatorname{diag}(\mathbf S\mathbf B\mathbf c^{(k)})$.
This normalization keeps the eigenvalues in \([0,1]\), so repeated powers remain bounded.
The filter is evaluated through repeated matrix--vector products, with $P+1$ coefficients learned for each mode in each layer.
In total, each layer has $2S_1+2S_2+K(P+1)+3$ trainable parameters.

\section{Experimental Results}
\label{sec:exp}

We evaluate graph recovery on synthetic signals and signal completion on weather data.
The proposed unrolled networks use $K=3$, $J=8$, $S_1=3$, $S_2=20$, $P=3$, and trained using Adam.
We train these networks using random masks hiding 15\,\% of the available observations, minimizing the normalized squared reconstruction error on the hidden entries.

\subsection{Synthetic Graph Recovery}
\label{sec:synth}

\noindent\textbf{Setting}: Each dataset has $N=32$ channels and $M=32$ windows of $T=360$ hourly samples.
We generate two stochastic block model prototypes~\cite{holland_stochastic_1983} with the same two groups of 16 nodes.
Their intra- and inter-group edge probabilities are $(0.35,0.10)$ and $(0.10,0.33)$, respectively.
Each mode--window graph is a convex mixture of the prototypes, with mixing weights from $\{0.2p\}_{p=1}^{4}$.
The three modes have periods of 115, 24, and 12 h, with corresponding autoregressive amplitude decay times of 72, 48, and 12 h.
Their complex amplitudes have covariance proportional to $\exp(-4\mathbf L)$.
We add white noise at 10-dB SNR and mask 10--50\,\% of entries, either randomly or as one contiguous gap per channel.
We generate 25 independent datasets and use 8 / 7 / 10 for training / validation / testing.

Baselines include MVMD~\cite{urrehman_multivariate_2019} followed by either independent Kalofolias graph learning~\cite{kalofolias_how_2016} or per-mode GraphDict~\cite{cappelletti_graphdictionary_2026}, and TVGMD~\cite{urrehman_timevarying_2023}.
To evaluate the benefit of algorithm unrolling, we also report the performance of the proposed alternating optimization solved iteratively.
For each condition, we tune baseline methods and the proposed iterative solver on the validation datasets by minimizing the normalized signal reconstruction error.
For the proposed unrolled networks, we consider $R=1$ and $R=2$ prototypes to validate the effectiveness of using multiple prototypes.
The proposed networks ($R=1,2$) are trained for 800 steps on eight datasets.
Their center frequencies are initialized at periods of 115, 24, and 12 h.

We measure graph recovery performance by the relative edge-weight error $\|\hat{\mathbf w}-\mathbf w\|_2/\|\mathbf w\|_2$, after rescaling both vectors to sum to $N/2$.
Errors are averaged over mode--window graphs.

\begin{figure}[t]
  \centering
  \includegraphics[width=\columnwidth]{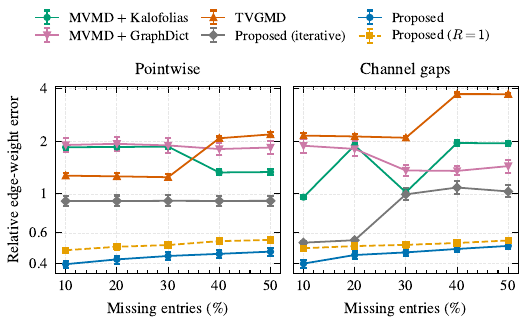}
  \caption{Graph error at 10-dB SNR, with $R=2$ for the proposed network. Means over ten test sets with 95\,\% confidence intervals.}
  \label{fig:sweep_graph}
\end{figure}

\noindent\textbf{Results}: Figure~\ref{fig:sweep_graph} shows that the proposed network ($R=2$) achieves the lowest mean graph error across all ten conditions.
Compared with $R=1$, it reduces mean error by 13--17\,\% for pointwise missing data and 7--18\,\% for channel gaps.
These results support using multiple prototypes when the graphs contain distinct shared patterns.

\subsection{Weather Completion}
\label{sec:real}

\input{tables/tab_amedas_nmse_v5_onecol}

\noindent\textbf{Setting}:
We use hourly Japan Meteorological Agency observations\footnote{JMA AMeDAS, \url{https://www.jma.go.jp}} from 145 stations over 2019--2025.
The data are organized into 168 nonoverlapping windows of 336 h each.
We evaluate signal completion using temperature data and examine the physical interpretability of the recovered modes using sea-level pressure data.

We randomly generate evaluation masks within each observation window to define five temperature completion tasks:
1) Pointwise: 20\,\% of valid entries are randomly hidden.
2) Station-window: all observations from 20\,\% of stations are hidden throughout the window.
3)--5) Gap completion: contiguous gaps of 6, 24, or 72 h are placed at random, hiding approximately 20\,\% of valid entries in each task.
The masked values are used only for evaluation and are excluded from preprocessing and model fitting.
To remove station-specific levels and slow drift, we linearly detrend each station within each window.
When observations are insufficient to fit a trend, we estimate it by inverse-square-distance averaging of trends fitted at up to eight nearby stations.
The residuals are normalized by a common scale, and missing entries are initialized from nearby stations.

We compare temporal linear interpolation, geographic Tikhonov regularization, IGNNK~\cite{wu_inductive_2021}, SAITS~\cite{du_saits_2023}, and IDGL~\cite{batreddy_inpainting_2025}.
For geographic Tikhonov regularization, we select the regularization weight from seven candidates by minimizing validation reconstruction error.
The proposed network uses $R=3$ and center frequencies fixed at periods of 120, 24, and 12 h.
We evaluate three ablations targeting graph filtering, structure sharing, and prototype diversity. 
First, we remove graph filtering from the mode updates. 
Second, we learn each mode–window graph independently. 
Third, we use a single prototype (\(R=1\)), so all graphs share one connectivity pattern up to scaling.

We also evaluate a variant of the proposed method that estimates the center frequencies jointly with the modes. 
Their initial values are drawn once per fit, uniformly at random from $[1/T,\,1/2 - 1/T]$ cycles per sample.
This comparison examines whether the network can adapt the center frequencies to the observed signals without prior knowledge of the periods.

The proposed networks are trained for 3,200 steps.
For self-supervised training of the proposed network, we additionally mask a subset of these input observations and minimize their reconstruction error.
Validation masks follow the same missing pattern as the evaluation task.
We report NMSE over evaluation entries across all windows, averaged over 3--5 evaluation masks.
For the proposed model, the across-mask standard deviation after seed averaging is below 4\,\% of the mean in all five tasks.

\noindent\textbf{Results}:
Table~\ref{tab:amedas} reports the NMSE for the five temperature completion tasks.
The proposed network ranks first or second in mean NMSE across all five tasks.
It achieves lower NMSE than SAITS on all three gap completion tasks, using only 488 trainable parameters compared with $\sim$730K for SAITS.

The ablations clarify how graph structure contributes to completion.
Removing graph filtering increases NMSE in all five tasks, supporting the use of spatial relationships alongside temporal information.
Learning each graph independently also gives higher NMSE than the proposed model for station-window task and all three gap completion tasks.
These results suggest that sharing connectivity patterns across modes and windows helps recover signals when observations are missing over entire stations or continuous intervals.
The gains from multiple prototypes are modest and task-dependent: \(R=1\) achieves lower mean NMSE for 24-h gaps, whereas \(R=3\) performs better for 72-h gaps. 
Stronger regularization with a single prototype may explain its advantage in some settings.

We also examine whether the recovered modes reflect physically meaningful patterns.
For temperature, the free-frequency variant recovers a mode whose spectral peak lies within one DFT bin of the 24-h line in all 30 fits under pointwise and station-window tasks.
This indicates that the network finds the daily component without prior knowledge of its period.

For sea-level pressure, we examine the spatial amplitude pattern of the 12-h component using the same stations and windows, with one pointwise mask and three seeds.
Station amplitudes are summarized by RMS over windows and averaged over seeds.
Across the 145 stations, they correlate with the expected latitude dependence $\cos^3(\mathrm{latitude})$~\cite{dai_diurnal_1999} ($r=0.827$) and with 12-h amplitudes fitted directly to the observations ($r=0.945$).
These results suggest that the component captures a physically meaningful spatial pattern.

\section{Conclusion}

We proposed joint signal restoration and multifold graph learning, representing component- and window-specific graphs as nonnegative combinations of shared prototypes.
We developed an alternating algorithm and unrolled it into a self-supervised neural network.
Synthetic experiments demonstrated improved graph estimation accuracy over the compared methods.
On weather data, the network achieved competitive completion performance with substantially fewer trainable parameters than a self-attention-based model.

\clearpage
\bibliographystyle{IEEEbib}
\bibliography{refs_revised_v8}

\end{document}

%% file: tables/tab_amedas_nmse_v5_onecol.tex
\begin{table}[t]
\centering
\caption{Temperature completion NMSE (bold: best, underlined: second, before rounding). Pt.: pointwise. Stn-win: station-window. Params.: number of trainable parameters.}
\label{tab:amedas}
\setlength{\tabcolsep}{1.5pt}
\begin{tabular*}{\columnwidth}{@{\extracolsep{\fill}}lrccccc@{}}
\toprule
& & & & \multicolumn{3}{c}{Gap length} \\
\cmidrule(lr){5-7}
Method / Task & Params. & Pt. & Stn-win & 6 h & 24 h & 72 h \\
\midrule
Linear interp. & 0 & 0.0317 & N/A & 0.1445 & 1.0668 & 1.1640 \\
Geo. Tikhonov & 0 & 0.1843 & 0.3150 & 0.1887 & 0.1990 & 0.2330 \\
IGNNK~\cite{wu_inductive_2021} & 386,536 & 0.2571 & 0.3386 & 0.2776 & 0.2802 & 0.3068 \\
SAITS~\cite{du_saits_2023} & 730,838 & 0.0864 & \textbf{0.2573} & 0.1080 & 0.1334 & 0.1747 \\
IDGL~\cite{batreddy_inpainting_2025} & 2 & 0.0381 & 0.4199 & 0.1673 & 0.2761 & 0.3317 \\
\midrule
Prop. (temporal) & 72 & 0.0301 & 0.3205 & 0.0758 & 0.1592 & 0.2009 \\
Prop. (indep.) & 328 & 0.0279 & 0.3153 & \underline{0.0635} & 0.1152 & 0.1540 \\
Prop. ($R=1$) & 488 & \underline{0.0278} & 0.2631 & 0.0637 & \textbf{0.1042} & \underline{0.1398} \\
Prop. (free) & 488 & 0.0286 & 0.2660 & 0.0640 & 0.1122 & 0.1584 \\
Proposed & 488 & \textbf{0.0278} & \underline{0.2612} & \textbf{0.0617} & \underline{0.1070} & \textbf{0.1365} \\
\bottomrule
\end{tabular*}
\end{table}